\documentclass[twocolumn]{jpsj2} 
\def\stacksymbols #1#2#3#4{\def\theguybelow{#2}
    \def\verticalposition{\lower#3pt}
    \def\spacingwithinsymbol{\baselineskip0pt\lineskip#4pt}
    \mathrel{\mathpalette\intermediary#1}}
\def\intermediary#1#2{\verticalposition\vbox{\spacingwithinsymbol
      \everycr={}\tabskip0pt
      \halign{$\mathsurround0pt#1\hfil##\hfil$\crcr#2\crcr
               \theguybelow\crcr}}}
\def\lapproxeq{\stacksymbols{<}{\sim}{2.5}{.2}}
\def\gapproxeq{\stacksymbols{>}{\sim}{3}{.5}}
\title{
Correlation Exponent and Anomalously Localized States at the
Critical Point of the Anderson Transition}
\author{H. Obuse and K. Yakubo}

\inst{Department of Applied Physics,
Graduate School of Engineering, Hokkaido University, Sapporo
060-8628, Japan.}

\abst{ We study the box-measure correlation function of quantum
states at the Anderson transition point with taking care of
anomalously localized states (ALS). By eliminating ALS from the
ensemble of critical wavefunctions, we confirm, for the first
time, the scaling relation $z(q)=d+2\tau(q)-\tau(2q)$ for a wide
range of $q$, where $q$ is the order of box-measure moments and
$z(q)$ and $\tau(q)$ are the correlation and the mass exponents,
respectively. The influence of ALS to the calculation of $z(q)$ is
also discussed. }

\kword{Anderson transition, critical states, multifractal,
anomalously localized states, two-dimensional symplectic system,
forced oscillator method}

\begin{document}

\maketitle

\section{Introduction}
\label{sec:1}

It is widely accepted that critical wavefunctions at the Anderson
metal-insulator transition are multifractal.\cite{Aoki1,Wegner1}
Since the multifractal nature at criticality forms a basis of the
single-parameter scaling argument of the Anderson
transition,\cite{Abrahams1} properties of critical wavefunctions
have been extensively studied.\cite{Nakayama1} Multifractal
properties of wavefunctions are usually characterized by the mass
exponent $\tau(q)$ or the multifractal spectrum $f(\alpha)$, where
$\tau(q)$ describes the system-size dependence of the local
moments $Z_q$ of the wavefunction distribution and $f(\alpha)$ is
the fractal dimension of the spatial distribution of boxes with
the Lipschiz-H\"older exponent $\alpha$. It is well known that
$\tau(q)$ is related to $f(\alpha)$ through the Legendre
transform. This relation provides a key criterion in judging
whether numerically calculated exponents are appropriate. However,
difficulty in calculating $f(\alpha)$ from its definition prevents
us from utilizing actually this relation for checking the accuracy
of the exponent. For this purpose, we should, thus, provide other
relations between exponents characterizing multifractality of a
critical wavefunction. A possible candidate is the scaling
relation $z(q)=d+2\tau(q)-\tau(2q)$, where $z(q)$ is the
correlation exponent describing the box-measure correlation
function. The box-measure correlation function is a generalization
of the local moment $Z_q$ and can be easily obtained in numerical
calculations. Although Pracz {\it et al.} have tried to confirm
this scaling relation for the quantum Hall
transition,\cite{Pracz1} $z(q)$ coincides with
$2+2\tau(q)-\tau(2q)$ only in a narrow range of $q$ around $q=0$
in their result. Since the profile of $z(q)$ near $q=0$ is not
sensitive to details of the multifractal distribution, it is
important to confirm the scaling relation for large $q$ values.

In order to clarify the reason why the calculated correlation
exponent $z(q)$ deviates from $2+2\tau(q)-\tau(2q)$ at large $q$'s
in their work, we notice the fact that some states are spatially
"localized" due to statistical fluctuations of disorder even not in
an insulating phase. These quantum states are called anomalously
localized states (ALS).
\cite{Altshuler1,Mirlin1,Muzykantskii1,Falko1,Smolyarenko1,Uski1,Nikolic1,Nikolic2,Uski2}
Major interest in ALS is focused on them in a metallic phase
because of the lack of theoretical tools for critical states.
There should, however, exist ALS also at the critical point,
because the statistical fluctuation of disorder at criticality
easily yields a pre-localized state compared to ALS in the metallic
regime. In fact, several previous works directly and indirectly
support the existence of ALS at the critical
point.\cite{Canali1,Steiner1,Mirlin2} Recently, we have shown that
the ratio of the number of ALS at criticality to the number of the
total states increases with system size $L$ and saturates at a
certain value for $L \rightarrow \infty$, while typical states are
kept to be multifractal.\cite{Obuse1} Since the scaling theory and
the renormalization group theory are based on the multifractal
nature of critical wavefunctions (i.e., divergence of the
correlation length at the transition point), conclusions
concerning properties just at the critical point obtained from
these theories are valid only for typical states. Therefore, the
scaling relation on the correlation exponent $z(q)$ should be
considered for typical critical states. This would be a primary
reason why the scaling relation has not been provided for large
$q$'s in the previous numerical work.

In this paper, we investigate the scaling relation for the
correlation exponent $z(q)$ by taking into account the influence
of ALS at the critical point. Critical wavefunctions are prepared
in two-dimensional symplectic systems which are described by the
SU(2) model.\cite{Asada1} In order to obtain $z(q)$ for {\it
typical} critical states, we compose an ensemble of critical
wavefunctions from which ALS are eliminated. The exponent $z(q)$
calculated for such an ensemble satisfies the scaling relation for
large $q$ values. We quantify how well the ensemble from which ALS
are eliminated describes statistical properties of typical
critical wavefunctions by a parameter $\Gamma^*$. It is found that
the range of $q$ for which $z(q)$ satisfies the scaling relation
depends on $\Gamma^*$. This paper is organized as follows. In
\S \ref{sec:2}, we give a quantitative definition of ALS at the
critical point based on the idea that ALS at criticality do not
show multifractality. In \S \ref{sec:3}, the scaling relation
for the correlation exponent is reminded in the scaling argument
for the box-measure correlation function. We briefly explain, in
\S \ref{sec:4}, the SU(2) model and a numerical method to obtain
eigenstates of this model. Results of our numerical study are
given in \S \ref{sec:5}. The insensitive nature of $z(q)$ near
$q=0$ to ALS is also discussed in this section. Section
\ref{sec:6} is devoted to conclusions.

\section{Definition of Anomalously Localized States}
\label{sec:2}

In order to study the influence of ALS quantitatively, we employ a
definition of ALS proposed in ref.~\citen{Obuse1}. This definition
is based on the idea that ALS are not multifractal as a
consequence of their localized nature. At first, we introduce the
box-measure correlation function $G_q(l,L,r)$ defined
by\cite{Cates1,Janssen2}
\begin{equation}
G_q (l,L,r) = \frac{1}{N_b N_{b_r}} \sum_b \sum_{b_r} \mu_{b(l)}^q \mu_{b_{r}(l)}^q ,
\label{eq:1}
\end{equation}
where $\mu_{b(l)}=\sum_{i \in b (l)} |\psi_i|^2$ and
$\mu_{b_r(l)}=\sum_{i \in b_r (l)} |\psi_i|^2$ are the box measures
for wavefunction amplitudes $\psi_i$, in a box $b(l)$ of size $l$
and in a box $b_r(l)$ of size $l$ fixed distance $r-l$ away from
the box $b(l)$, respectively. $N_b$ (or $N_{b_r}$) is the number
of boxes $b(l)$ [or $b_r(l)$], and the summation $\sum_{b}$ (or
$\sum_{b_r}$) is taken over all boxes $b(l)$ [or $b_r(l)$] in the
system of size $L$. If a wavefunction is multifractal,
$G_q(l,L,r)$ should behave as \cite{Janssen2}
\begin{equation}
G_q(l,L,r) \propto l^{x(q)} L^{-y(q)} r^{-z(q)},
\label{eq:2}
\end{equation}
where $x(q)$, $y(q)$, and $z(q)$ are exponents describing
multifractal correlations of the amplitude distribution. This
relation is sensitive to ALS as demonstrated in ref.~18 and then
suitable for defining ALS. To find the $l$ and $r$ dependences of
$G_q(l,L,r)$, we concentrate on the following functions,
\begin{equation}
Q_q(l) = G_q(l,L,r=l) \propto l^{x(q)-z(q)},
\label{eq:3}
\end{equation}
and
\begin{equation}
R_q(r) = G_q(l=1,L,r) \propto r^{-z(q)}.
\label{eq:4}
\end{equation}
In order to quantify non-multifractality of a specific
wavefunction, it is convenient to introduce variances $\text{Var}
(\log Q_2)$ and $\text{Var} (\log R_2)$ from the linear functions
of $\log l$ and $\log r$, $\log Q_{2}(l) = [x(2)-z(2)] \log l +
c_Q$ and $\log R_{2}(r) = -z(2) \log r + c_R$, respectively,
calculated by the least-square fit. From these variances, a
quantity $\Gamma$ is defined by
\begin{equation}
\Gamma(L,\lambda) = \lambda \text{Var}(\log Q_2) + \text{Var} (\log R_2),
\label{eq:5}
\end{equation}
where $\lambda$ is a factor to compensate the difference between
average values of $\text{Var}(\log Q_2)$ and $\text{Var}(\log
R_2)$. Using $\Gamma$ given by eq.~(\ref{eq:5}), the quantitative
and expediential definition of ALS at criticality is presented by
\begin{equation}
\Gamma > \Gamma^*,
\label{eq:6}
\end{equation}
where $\Gamma^*$ is a criterial value of $\Gamma$ to distinguish
ALS from multifractal states. We can compose a {\it refined
ensemble} by eliminating ALS from a set of critical wavefunctions.
The quality of the refined ensemble is controlled by $\Gamma^*$.

\section{Scaling Relation for the Correlation Exponent}
\label{sec:3}

In this section, we give a brief explanation of the scaling
relations between the exponents $x(q)$, $y(q)$, $z(q)$, and
$\tau(q)$. The mass exponent $\tau(q)$ which is commonly used in
the multifractal analysis is defined by
\begin{equation}
Z_q (l) \equiv \sum_b \mu_{b (l)}^q \propto l^{\tau(q)}.
\label{eq:7}
\end{equation}
The local moment $Z_q(l)$ is related to the correlation function
$G_q(l,L,r)$. Comparing definitions of $G_q$ and $Z_q$
[eqs.~(\ref{eq:1}) and (\ref{eq:7})], we have
\begin{equation}
G_q(l,L,r=l) = \frac{Z_{2q}(l)}{N_b}.
\label{eq:8}
\end{equation}
Therefore, the relation $N_b = (L/l)^d$
and eqs.~(\ref{eq:2}) and (\ref{eq:7}) lead
\begin{equation}
x(q)-z(q) = y(q) = d+\tau(2q).
\label{eq:9}
\end{equation}
For the case of $r=L$, eq.~(\ref{eq:1}) gives
\begin{equation}
G_q(l,L,r=L) = \frac{1}{N_b^2} \left ( \sum_b \mu_{b(l)}^q \right)^2.
\label{eq:10}
\end{equation}
Since $G_q(l,L,r=L) \propto l^{x(q)} L^{-y(q)-z(q)}$ from
eq.~(\ref{eq:2}), the exponents are related as
\begin{equation}
x(q)=y(q)+z(q)=2d+2\tau(q).
\label{eq:11}
\end{equation}
From eqs.~(\ref{eq:9}) and (\ref{eq:11}), we obtain
\begin{eqnarray}
x(q) \hspace{-3.2mm} &=& \hspace{-3.2mm} 2d + 2\tau(q), \label{eq:12} \\
y(q) \hspace{-3.2mm} &=& \hspace{-3.2mm} d +\tau(2q),  \label{eq:13}
\end{eqnarray}
and for the correlation exponent
\begin{equation}
z(q) = d+ 2\tau(q) - \tau(2q). \label{eq:14}
\end{equation}
The last relation with $q=1$ is equivalent to the well-know
scaling relation $\eta=d-D_2$,\cite{Chalker1} where $\eta$ is the
exponent describing the two-particle correlation function
$S(r)=\rho \langle |\psi(0)|^2 |\psi(r)|^2 \rangle$, $\rho$ is the
density of states, and $D_2$ is the correlation dimension of
critical wavefunctions. This equivalency can be understood from
the relations $z(1)=\eta$ because of $G_1(l=1,L,r)\propto S(r)$,
$\tau(1)=0$, and $\tau(2)=D_2$.

The above relations (\ref{eq:12})-(\ref{eq:14}) hold for any $q$
in principle if the wavefunction is multifractal. However, the
previous work confirming the scaling relation eq.~(\ref{eq:14})
for the quantum Hall transition\cite{Pracz1} shows that the
correlation exponent $z(q)$ coincides with $2+2\tau(q)-\tau(2q)$
for a narrow range of $q$ $(|q|<1.5)$. As mentioned in
\S \ref{sec:1}, we consider that the disagreement at large
$|q|$'s is a consequence of ALS. It should be noted that the
agreement for small $|q|$'s is rather trivial. Even if the
wavefunction is not multifractal, the exponents satisfy the
scaling relation eqs.~(\ref{eq:12})-(\ref{eq:14}) for $q=0$,
because $\tau(0)=-d$ and $x(0)=y(0)=z(0)=0$ for any $\mu_{b(l)}$.
Increasing $|q|$ slightly from zero, the correlation exponent
$z(q)$, for example, is well approximated by the parabolic form
\begin{equation}
z(q)=2(\alpha_0-d) q^2,
\label{eq:15}
\end{equation}
where $\alpha_0$ is the Lipschiz-H\"older exponent giving the
maximum of $f(\alpha)$. If $\alpha_0$ is insensitive to details of
the distribution of measures, $z(q)$ approximately satisfies
eq.~(\ref{eq:14}) near $q=0$, because $\tau(q)$ is given by
$(1-q)[q(\alpha_0-d)-d]$ in the parabolic approximation. We
should, thus, confirm the scaling relations for large $|q|$'s.
Since the exponent $z(q)$ characterizes directly the spatial
correlation of wavefunction amplitudes, we focus only on the
relation (\ref{eq:14}) hereafter.

\section{System and Numerical Method}
\label{sec:4}

Considering the advantage of system sizes, we focus our attention
on the Anderson transition in two-dimensional electron systems
with strong spin-orbit interactions, in which systems have no
spin-rotational symmetry but have the time-reversal one.
Hamiltonians describing these systems belong to the symplectic
ensemble. Among several models belonging to this universality
class, we adopt the SU(2) model because of its small scaling
corrections. The Hamiltonian of the SU(2) model\cite{Asada1} is
given by
\begin{equation}
\boldsymbol{H} = \sum_{i} \varepsilon_i  \boldsymbol{c}_{i}^\dagger  \boldsymbol{c}_{i}
  -  V \sum_{i,j} \boldsymbol{R}_{ij} \boldsymbol{c}_{i}^\dagger \boldsymbol{c}_{j},
\label{eq:16}
\end{equation}
where $ \boldsymbol{c}_{i}^\dagger$ ($ \boldsymbol{c}_{i}$) is the
creation (annihilation) operator acting on a quaternion state
vector, $\boldsymbol{R}_{ij}$ is the quaternion-real hopping
matrix element between the sites $i$ and $j$, and $\varepsilon_i$
denotes the on-site random potential distributed uniformly in the
interval $[-W/2, W/2]$. (Bold symbols represent quaternion-real
quantities.)
The matrix element
$\boldsymbol{R}_{ij}$ is given by
\begin{eqnarray}
\boldsymbol{R}_{ij} &=& \cos \alpha_{ij} \cos \beta_{ij} \boldsymbol{\tau}^0
+ \sin \gamma_{ij} \sin \beta_{ij}\boldsymbol{\tau}^1
\nonumber
\\
&-& \cos \gamma_{ij} \sin \beta_{ij}\boldsymbol{\tau}^2
+ \sin \alpha_{ij} \cos \beta_{ij}\boldsymbol{\tau}^3,
\label{eq:17}
\end{eqnarray}
for the nearest neighbor sites $i$ and $j$, and
$\boldsymbol{R}_{ij}=0$ for otherwise. Here,
$\boldsymbol{\tau}^\mu$ ($\mu=0,1,2,3$) is the primitive element
of quaternions.\cite{Kyrala1} Random quantities $\alpha_{ij}$ and
$\gamma_{ij}$ are distributed uniformly in the range of
$[0,2\pi)$, and $\beta_{ij}$ is distributed according to the
probability density $P(\beta) d \beta = \sin(2\beta) d \beta$ for
$0 \le \beta \le \pi/2$. Randomly distritbuted hopping matrix
elements shorten the spin relaxation length which is a dominant
irrelevant length scale, and scaling corrections can be negligible
in the SU(2) model. It is known that the localization length
exponent $\nu$ of this model is $2.73 \pm 0.02$ and the critical
disorder $W_c$ is $5.952V$ at $E=1.0V$.\cite{Asada1}

Critical wavefunctions of the SU(2) model have been calculated by
using the forced oscillator method (FOM)\cite{Nakayama2} extended
to the eigenvalue problem of quaternion-real matrices. Of course,
the Hamiltonian eq.~(\ref{eq:16}) can be represented by complex
numbers, and we can use the usual FOM for complex Hermitian
matrices to solve the eigenvalue problem. The modified FOM for
quaternion-real matrices, however, enables us to calculate
eigenvalues and eigenvectors within about a half of CPU
time.\cite{Obuse2} It should be remarked that the obtained
eigenvector is a quaternion-real vector. This vector represents
two physical states simultaneously, which correspond to the
Kramers doublet. Since the amplitude distribution of these
degenerate states are the same, we analyze one of the calculated
Kramers doublet.

\section{Results}
\label{sec:5}

In order to study the correlation exponent $z(q)$ for critical
wavefunctions of the SU(2) model, we calculate $10^4$ critical
wavefunctions of this model at $E=1.0V$ and $W=5.952V$ by the FOM.
Each eigenstate is obtained for a single disorder realization.
Periodic boundary conditions are imposed in the $x$ and $y$
directions in systems of size $L=120$.

\begin{figure}[t]
\begin{center}
\includegraphics[width=7.5cm]{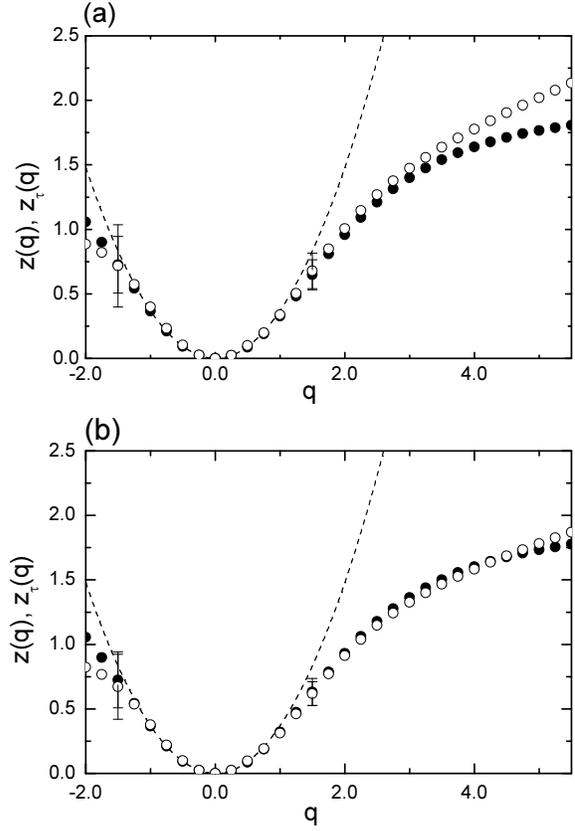}
\end{center}
\caption{ Calculated correlation exponent $z(q)$ (open circles)
and the right-hand side of eq.~(\ref{eq:14}) $[z_\tau
(q)=d+2\tau(q)-\tau(2q)]$ (filled circles). Figures labelled by
(a) and (b) correspond to the results for the original ensemble and the
refined ensemble, respectively. The dashed lines show the parabolic
approximation of $z(q)$ given by eq.~(\ref{eq:15}) with $\alpha_0=2.18$. }
\label{fig:1}
\end{figure}

Figure \ref{fig:1}(a) shows the $q$ dependence of the correlation
exponent $z(q)$ and the right-hand side of eq.~(\ref{eq:14}), 
i.e., $d+2\tau(q)-\tau(2q) [\equiv z_\tau(q)]$, calculated for
the whole set of the prepared critical wavefunctions (the original
ensemble). The exponents $z(q)$ and $\tau(q)$ are calculated from
the geometric means of $R_q(r)$ and $Z_q(l)$ by the least-square
fit, respectively. It is found that $z(q)$ deviates from
$z_\tau(q)$ at large $|q|$'s. The degree of agreement is similar
to the result of the previous work for the quantum Hall
transition.\cite{Pracz1} We see that $z(q)$ for $q \gapproxeq 5$
exceeds the value of $2$ which is the theoretical bound of $z(q)$.

It should be noted that the scaling relation eq.~(\ref{eq:14}) is
valid for {\it typical} critical wavefunctions. As shown in our
previous work,\cite{Obuse1} the distribution function of $\Gamma$
defined by eq.~(\ref{eq:5}) has its peak at $\Gamma=0$, which
implies that typical critical wavefunctions are multifractal.
Since the geometric mean usually represents the typical value
(mode value), it seems that the geometric mean $\langle R_q(r)
\rangle$ used in our calculations gives the correct exponent
$z(q)$ of multifractal critical wavefunctions without influence
from ALS. However, the distribution function of $\Gamma$ defined
for the correlation function $G_q$ becomes broad with increasing
$|q|$, and then the geometric mean could not represent the typical
value for an ensemble composed of an insufficient number of
elements due to enhanced statistical fluctuations. In fact, we
have confirmed that the correlation exponent calculated from the
typical profile of $R_q(r)$ largely deviates from the exponent
computed from $\langle R_q(r) \rangle$ for large $|q|$. Although
the correlation exponent appeared in eq.~(\ref{eq:14}) should be
computed from the typical profile of $R_q(r)$ $[R_q^{\text{typ}}
(r)]$, it is actually quite difficult in numerical calculations to
calculate precise value of $z(q)$ from $R_q^{\text{typ}} (r)$.

In this paper, we show that the elimination of ALS from the
original ensemble of critical wavefunctions is efficient technique
to obtain accurate value of $z(q)$ even for large $|q|$.
At first, we compose a refined ensemble by eliminating ALS from
the original ensemble. The refined ensemble is characterized by
the parameters $\lambda=3$ and $\Gamma^*=0.06$, and contains
$6962$ critical wavefunctions. The correlation exponent calculated
from $\langle R_q(r) \rangle$ for this refined ensemble is
presented in Fig.~\ref{fig:1}(b). The exponent $z(q)$ is in
agreement with $z_\tau(q)$ for a wider range of $q$, and does not
exceed the theoretical bound $2$. This demonstrates the efficiency
of the elimination scheme of ALS to calculate precise exponents
defined at the critical point. We find, however, that the exponent
$z(q)$ deviates from $z_\tau(q)$ for $q \lapproxeq -2$ even for
the refined ensemble. What is even worse, $z(q)$ becomes negative
for $q \ll -2$. To see the reason of this unreasonable result, let
us consider a simple case of $R_q(r)$, namely, $R_q(r)$ for
quantities $\mu_i$ distributed uniformly in the range of $[0,1]$
without any spatial correlations. From the definition of $R_q(r)$
[or $G_q(l,L,r)$], we have
\begin{equation}
R_q(r)=\frac{1}{N N_r}\sum_i \sum_{j \in r(i)} \mu_i^q \mu_j^q,
\label{eq:19}
\end{equation}
where $N$ is the total site number, $\sum_{j \in r(i)}$ represents
the summation over sites $j$ away from the site $i$ by a distance
$r$, and $N_r$ is the number of such sites.
The uncorrelated distribution of $\mu_i$ makes it possible to
replace $\mu_i^q$ by its average value $\overline{\mu^q}$, and we
obtain
\begin{equation}
R_q(r)=\frac{(\overline{\mu^q})^2}{N N_r} \sum_i \sum_{j \in r(i)} = (\overline{\mu^q})^2.
\label{eq:20}
\end{equation}
From this relation, it seems that the exponent $z(q)$ becomes zero
for any $q$, because $\overline{\mu^q}$ does not depend on $r$.
This is, however, not true for negative $q$. The distribution
function of $\mu^q$ is given by $\theta(t)/[qt^{(q-1)/q}]$ for
positive $q$ and $[1-\theta(t)]/[|q|t^{(q-1)/q}]$ for negative
$q$, where $t=\mu^q$ and $\theta(t)$ is the step function defined
as $\theta(t)=1$ for $0 \le t \le 1$ and $\theta(t)=0$ for $t>1$.
For $q>0$, the distribution is truncated at $t=1$ and exponent
$(q-1)/q$ in the distribution function is less than unity, which
gives a finite average value of $\mu^q$. On the contrary, the
distribution function for negative $q$ extends over the range of
$[1,\infty)$. The exponent $(q-1)/q$ becomes less than $2$ for
$q<-1$, which leads an infinite average value of $\mu^q$. Thus we
cannot define the exponent $z(q)$ from eq.~(\ref{eq:20}) for
$q<-1$. The divergence of $\overline{\mu^q}$ is a consequence that
the distribution function of $\mu$ is finite at $\mu=0$. Since a
distribution of multifractal measures is not uniform (but
log-normal approximately) and spatial correlations of them must be
taken into account, the above argument cannot be applied
straightforwardly to refined ensembles of critical wavefunctions.
Nevertheless, the log-normal distribution function taking
relatively large values near $\mu=0$ induces a numerical
instability for $q<0$. This instability leads the unreasonable
result for $q \ll -2$.
\begin{figure}[t]
\begin{center}
\includegraphics[width=8cm]{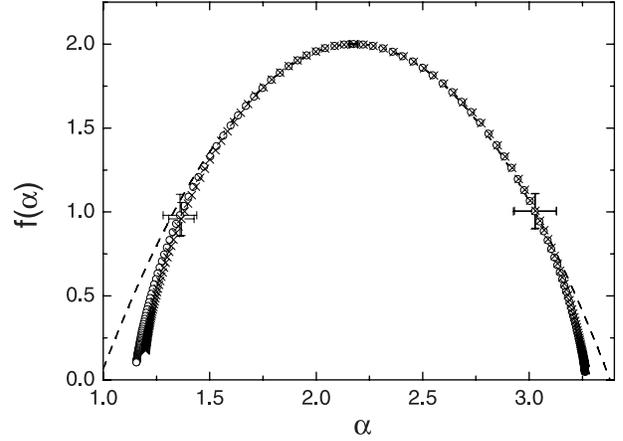}
\end{center}
\caption{ Multifractal spectra $f(\alpha)$ for the original
ensemble (circles) and the refined ensemble (crosses). Crossed
error bars indicate representative values of errors in $f(\alpha)$
and $\alpha$. The dashed line indicates the parabolic approximation
eq.~(\ref{eq:18}) with $\alpha_0=2.18$.} \label{fig:2}
\end{figure}

We see from Fig.~\ref{fig:1} that the exponent $z_\tau (q)$ is not
sensitive to ALS. This is consistent with the fact that the local
moment $Z_q(l)$ is not enough to distinguish ALS from multifractal
wavefunctions. The multifractal spectra calculated for the
original and the refined ensembles also indicate this fact as
shown in Fig.~\ref{fig:2}, where $f(\alpha)$ has been calculated
by the $q$-microscope.\cite{Chhabra1} Especially, these two
spectra almost collapse in the vicinity of $\alpha_0$ giving the
maximum of $f(\alpha)$, which corresponds to $q=0$. Since the
profile of $f(\alpha)$ close to $\alpha=\alpha_0$ can be
approximated by the parabolic form
\begin{equation}
f(\alpha)=2-\frac{(\alpha_0-\alpha)^2}{4(\alpha_0-2)},
\label{eq:18}
\end{equation}
this implies that $\alpha_0$ is hardly affected by ALS. (Dashed
line in Fig.~\ref{fig:2} shows this parabolic approximation with
$\alpha_0=2.18$.) Using this value of $\alpha_0$, we can draw the
parabolic approximation of $z(q)$ given by eq.~(\ref{eq:15}) as
shown in Fig.~\ref{fig:1}. The insensitivity of $\alpha_0$ to ALS
leads insensitive $z(q)$ to ALS near $q = 0$. In fact, $z(q)$
shown in Fig.~\ref{fig:1} coincides with $z_\tau(q)$ and the
parabolic approximation in the vicinity of $q=0$ even for the
original ensemble.

\begin{figure}[t]
\begin{center}
\includegraphics[width=8cm]{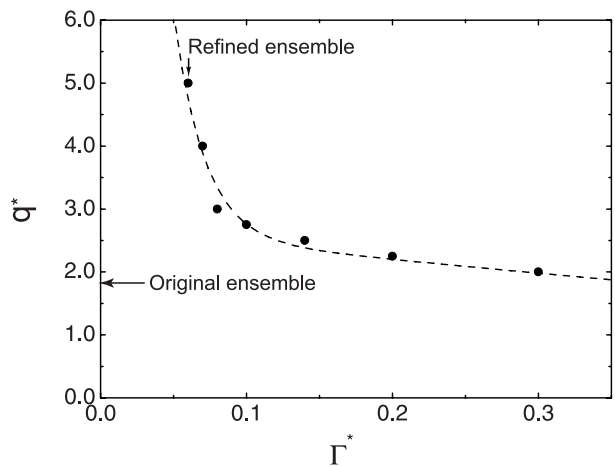}
\end{center}
\caption{
Quantity $q^*$ representing the quality of the refined ensemble
as a function of $\Gamma^*$.
$q^*$ is defined as $|z(q)-z_\tau(q)|>0.04$ for $q>q^*$.
Arrows indicate the data points corresponding to the refined ensemble
used in Fig.~\ref{fig:1} ( or \ref{fig:2}) and the original one.
}
\label{fig:3}
\end{figure}

Apparently, the {\it quality} of the refined ensemble depends on
the choice of the value of $\Gamma^*$ in eq.~(\ref{eq:6}).
Figure~\ref{fig:3} exhibits this quantitatively. The longitudinal
axis $q^*$ of this figure represents the value of $q$ $(>0)$ above
which $|z(q)-z_\tau(q)|$ is larger than $0.04$. The data points
indicated by arrows correspond to the original ensemble and the
refined ensemble used in Figs.~\ref{fig:1} and \ref{fig:2}. This
result indicates that the exponent $z(q)$ coincides with
$z_\tau(q)$ in the wider range of $q$ when the degree of refining the
ensemble becomes higher. Namely, the high quality refined ensemble
ensures the scaling relation eq.~(\ref{eq:16}) with large $q$. It
seems that $q^*$ becomes infinite when $\Gamma^*$ goes to zero.
The quantity $q^*$, however, does not diverge actually. Too small
$\Gamma^*$ makes the quality of the refined ensemble rather worse,
because the number of samples in the refined ensemble becomes
insufficient to provide a good statistics.

\section{Conclusions}
\label{sec:6}

We have investigated the box-measure correlation function
$G_q(l,L,r)$ of critical wavefunctions of the SU(2) model which
belongs to the two-dimensional symplectic class and confirmed the
scaling relation for the correlation exponent $z(q)$ in a wide
range of $q$. It is found that ALS prevent us from calculating
accurate values of the correlation exponent with large $|q|$'s. To
obtain $z(q)$, we eliminated ALS from the original ensemble of
critical wavefunctions and composed a refined ensemble for which
the geometric mean of the box-measure correlation functions
represents properly the typical profile of $G_q(l,L,r)$. It is
also found that the range of $q$ within which the calculated
$z(q)$ satisfies the scaling relation becomes wider when the
quality of the refined ensemble becomes better (namely, decreasing
$\Gamma^*$). We have elucidated that a direct calculation of the
exponent $z(q)$ encounters a numerical instability for $q \ll -2$,
which originates from the fact that the distribution function of
multifractal measure $\mu$ has relatively large values near
$\mu=0$. Some ingenious techniques in numerical calculations would
be required to obtain precise values of $z(q)$ at negative $q$'s.
In the present paper, the dependence of the correlation function
$G_q(l,L,r)$ on the box-box distance $r$ has been examined only
for $l=1$. If we can deal with huge systems, $G_q(l,L,r)$ with
$l>1$ would provide a more precise value of $z(q)$ for $q \ll -2$,
because the numerical instability at negative $q$'s might be
moderated due to an average effect within each box. In our
demonstrations by using the SU(2) model, the value of $q$ to
eliminate ALS has been fixed at $2$, while $G_q$ has been analyzed
for many $q$ values. Essentially, the choice of the value of the
moment $q$ for the elimination does not influence the quality of
the obtained refined ensembles. It is, however, better to use a
positive small value of $q$ for ensuring a numerical stability.

The composition of the refined ensemble
is efficient also for evaluating other properties at the critical
point than the scaling relation treated in this paper.
We emphasize that the scaling relation eq.~(\ref{eq:14})
can be used for judging how well the refined ensemble represents
typical critical wavefunctions. It is also interesting to study
statistical properties of the ensemble of ALS, namely, the
complementary set of the refined ensemble. Further investigations
of ALS provide us deep insight into the nature of criticality.

\begin{acknowledgements}
We are grateful to T. Nakayama and T. Ohtsuki for helpful
discussions. This work was supported in part by a Grant-in-Aid for
Scientific Research from Japan Society for the Promotion of
Science (No.~$14540317$). Numerical calculations in this work have
been mainly performed on the facilities of the Supercomputer
Center,Institute for Solid State Physics, University of Tokyo.
\end{acknowledgements}


\begin{thebibliography}{99}
\bibitem{Obuse2}
Details of the forced oscillator method for quaternion-real
matrices will be published elsewhere.
\end{thebibliography}
\end{document}